\documentclass[journal=jpctc,manuscript=article]{achemso}

\setkeys{acs}{articletitle = true}

\usepackage[utf8]{inputenc}

\usepackage{amsmath, amssymb}
\usepackage{tabularx,booktabs,array,dcolumn}
\usepackage{setspace}
\usepackage{vmargin}
\usepackage{graphicx,psfrag,subfigure}
\usepackage{color}

\usepackage[all]{xy}

\usepackage{enumitem}
\usepackage{mdframed}
\usepackage{xcolor}

\newcommand{\mx}[1]{\boldsymbol{#1}}
\newcommand{\bos}[1]{\boldsymbol{#1}}
\newcommand{\mr}[1]{\mathrm{#1}}
\newcommand{\dd}{\mathrm{d}}
\newcommand{\pd}[2]{\frac{\partial #1}{\partial #2}}

\newcommand{\rr}[1]{\mx{r}^{(#1)}}
\newcommand{\ri}[2]{r^{(#1)}_{#2}}

\def\ra0{\mx{a}}

\def\betam{\beta_M}
\def\omem{\omega_M}

\def\hti{\hat{H}}
\def\kbol{k_\text{B}}

\def\nns{K} 
\def\niter{I} 

\author{B\'ela Szekeres}
\affiliation{Institute of Chemistry, E\"otv\"os Lor\'and University, P\'azm\'any P\'eter s\'et\'any 1/A, Budapest, H-1117, Hungary}
\affiliation{Department of Numerical Analysis, Faculty of Informatics, E\"otv\"os Lor\'and University, P\'azm\'any P\'eter s\'et\'any 1/C, Budapest, H-1117, Hungary}

\author{L\'ivia B. P\'artay}
\email{l.bartokpartay@ reading.ac.uk}
\affiliation{Department of Chemistry, University of Reading, Whiteknights, Reading, RG6 6AD, UK}

\author{Edit M\'atyus}
\email{matyus@ chem.elte.hu}
\affiliation{Institute of Chemistry, E\"otv\"os Lor\'and University, P\'azm\'any P\'eter s\'et\'any 1/A, Budapest, H-1117, Hungary}

\title[Path-integral nested sampling]{%
Direct computation of the quantum partition function by path-integral nested sampling
}

\begin{document}


\begin{abstract}
\noindent %
In the present work we introduce a computational approach
to the absolute rovibrational quantum partition function using 
the path-integral formalism of quantum mechanics in combination with the nested sampling
technique. 
The numerical applicability of path-integral nested sampling is
demonstrated for small molecules of spectroscopic interest. 
The computational cost of the method is determined by the evaluation time of a point on the potential-energy surface (PES). 
For efficient PES implementations, the path-integral nested-sampling
method can be a viable alternative to the direct Boltzmann summation technique of variationally computed rovibrational energies, especially for medium-sized molecules and
at elevated temperatures.
\end{abstract}

\maketitle

%
%
\section{Introduction\label{ch:intro}}
\noindent%
In order to build useful databases for the modelling of radiative and chemical processes which
take place under exotic conditions in the laboratory, on Earth, in the interstellar space, or in the atmosphere of distant exoplanets, we need access to accurate 
quantum partition functions and thermodynamic data for polyatomic molecules over a wide temperature range \cite{H3p2017,SoSiHeYuHiTe14,WeChBo08,YuTeBaHoTi14}.
As Ref.~\cite{SoSiHeYuHiTe14} points out ``the [absolute] partition function is necessary 
to establish the correct temperature dependence of
spectral lines and their intensity''.

The quantum partition function of small molecular systems 
is usually calculated as the Boltzmann sum
of variationally computed rovibrational energies including appropriate factors, which account for both the degeneracies
and the spin-statistical weights. 
Although, the direct variational solution of the rovibrational Schr\"odinger equation \cite{LaNa02,YuThJe07,MaCzCs09,FaMaCs11}
provides all quantum dynamical information about the molecule, 
it is computationally feasible only for the smallest systems.
And even for these, converging the value of the partition function at elevated temperatures has 
been considered as a challenging task \cite{H3p2017,SoSiHeYuHiTe14,WeChBo08,YuTeBaHoTi14}, 
the variational computation of an increasing number of rovibrational energies is tedious,
and near or beyond the lowest dissociation threshold it is non-trivial \cite{SzCs15}.

In order to overcome this problem, approximations to this rigorous approach can be made, 
\emph{e.g.,} the quantum partition function is very often 
estimated within the harmonic-oscillator-rigid-rotor model. However, the relative error of 
the quantum partition function from this model is several percent \cite{JPCRD16} in an ideal case, 
moreover, this model is often qualitatively incorrect, \emph{e.g.,} for molecules with several torsional degrees of freedom.

Instead of looking for approximate models, 
we can settle for certain integral properties without explicitly computing the fully detailed
quantum dynamical information.
In this spirit, thermodynamic quantities of complex systems are usually calculated using 
the path-integral formalism of quantum mechanics \cite{FeHiSt10,Cermp95}.
For the theoretical description and understanding of many 
chemically interesting processes, we need only relative quantities, \emph{e.g.,}  free-energy differences, and 
there are efficient approaches for computing a variety of these.
Furthermore, corrections to simple model systems can be determined. For example, anharmonic corrections in larger systems (in which rotations are 
neglected or assumed to be separable) have been obtained using 
path-integral Monte Carlo techniques
and thermodynamic integration 
starting from uncoupled quantum harmonic oscillators as a reference 
\cite{MiCl03,MiCl04,ChPrBe06,AzEnDoMa11}. Later, not only anharmonicities
but also the rotation-vibration coupling was fully included in the 
path-integral computation of equilibrium isotope effects.\cite{ZiVa09}
%
In a series of papers \cite{ToTaTr92,ToTr92,ToZhLiTr93,MiTr03,LyMiTr04} 
Truhlar and co-workers elaborated on the Monte Carlo Fourier path integral method
which gives access to the absolute rotational-vibrational
partition function of molecules (with the full inclusion of anharmonicities and 
rotation-vibration coupling). The separable rotations approximation was found to be in
an error of $~7$~\% with respect to the exact result 
for small triatomic molecules \cite{ToZhLiTr93}. 
Most recently, further developments of this method made it possible 
to compute the rovibrational partition function of 
the methane molecule up to 3000~K.\cite{MiTr15,MiTr16}
Ref.~\cite{RiPrVa01} presented a path-integral
Monte Carlo computation of the quantum mechanical rovibrational partition function and
numerical results for the diatomic hydrogen molecule.
It was pointed out also in this work that path-integral based methods automatically include contributions from unbound states, which can be accounted for by the variational-Boltzmann summation technique only in a very approximate manner.

As to van-der-Waals complexes, in which contributions from unbound states are significant
already at moderate temperatures, 
a highly efficient path-integral Monte Carlo approach was developed 
for the second virial coefficient \cite{GaJaSzHa14}, which is related to the ratio of the dimer's 
to the monomers' partition function. Recent applications of this method
include the H$_2\cdot$CO dimer using a flexible spectroscopic PES \cite{GaJaSzHa17}.

In the present work we describe and present the first numerical applications
of path-integral nested sampling, 
a novel path-integral method which provides 
the absolute rovibrational quantum partition function 
value (including all anharmonic ``effects'' and rovibrational coupling) 
for molecular systems with an ``arbitrary'' connectivity.
Nested sampling \cite{Sk04,Sk06} has already been successfully used 
for a series of different classical systems \cite{PaBaCs10,PaBaCs14,BaPaBa16,BaBeSa17}
to compute absolute thermodynamic quantities 
without having to have any \emph{a priori} knowledge of 
the particular shape and location of the important basins in the configuration space.
In the numerical applications, we focus on a temperature range (above a particular value -- 
typically above 100~K) in which 
the effect of the spin-statistical weights (SSW) can 
be accounted for by a simple multiplication of the 
partition function computed without SSWs (more precisely, with equal SSWs) \cite{SoSiHeYuHiTe14,YuTeBaHoTi14}, which is 
straightforwardly obtained with path-integral nested sampling.

%
%
\section{Quantum Hamiltonian and partition function\label{ch:qpart}}
\noindent %
The quantum nuclear Hamiltonian of a molecule of $N$ atoms (in atomic units) is
\begin{align}
  \hat{H}'
  =
  -\sum_{i=1}^N \sum_{\alpha}
  \frac{1}{2m_i} \pd{^2}{r^2_{i\alpha}}
  + 
  V(|\mx{r}_i-\mx{r}_j|), 
  \label{eq:Hamcart}
\end{align}
where $m_i$ are the masses associated to the nuclei, $\alpha=1(x),2(y),3(z)$ 
labels the laboratory-frame (LF) axes, and 
$V(|\mx{r}_i-\mx{r}_j|)$ is the potential energy surface (PES), which depends on
the relative positions of the nuclei for an isolated system.
The energy-spectrum of $\hat{H}'$ is continuous due to the overall translation
of the system. In the present work, we separate off the overall translation:
subtract the kinetic energy of the center of mass ($\hat{T}_\text{CM}$) 
and introduce Jacobi Cartesian
coordinates, $x_{i\alpha}$ ($i=1,\ldots,N-1$).
We chose Jacobi coordinates because the corresponding Hamiltonian 
has a simple form, similar to Eq.~(\ref{eq:Hamcart}), 
without any derivative cross terms:
\begin{align}
  \hti
  &=
  \hat{H}'
  - 
  \hat{T}_\text{CM}
  \nonumber \\
  &=
  -\sum_{i=1}^{N-1} \sum_{\alpha}
  \frac{1}{2\mu_i} \pd{^2}{x^2_{i\alpha}}
  + 
  V(|\mx{r}_i-\mx{r}_j|)
  \label{eq:Hamjac}
\end{align}
only the $m_i$ physical masses are replaced with
the $\mu_i\ (i=1,\ldots,N-1)$ reduced masses
corresponding to the selected Jacobi coordinate set (see Appendix).
Eigenvalues of $\hti$ are the rotation-vibration energies, $E_i$, and 
the corresponding rotation-vibration partition function is 
\begin{align}
  Q(\beta)
  &=
  \text{Tr}\left[\text{e}^{-\beta\hti}\right]
  =
  \sum_{n=1}^{n_\text{max}}
    \text{e}^{-\beta E_i},
  \label{eq:qrovib}
\end{align}
where the last equality holds if only bound states are populated 
at temperature $T=1/(\kbol\beta)$ ($\kbol$\ is the Boltzmann constant).
Obviously, the partition function corresponding to the full, laboratory-fixed Hamiltonian, $\hat{H}'$, 
is related to the rovibrational partition function, Eq.~(\ref{eq:qrovib}), as
\begin{align}
  Q'(\beta)
  =
  \text{Tr}\left[\text{e}^{-\beta\hat{H}'}\right]
  =
  \text{Tr}\left[\text{e}^{-\beta(\hat{H}+\hat{T}_\text{CM})}\right]  
  =
  Q(\beta)\cdot Q_\text{CM}(\beta),
  \label{eq:Qticm}
\end{align}
where $Q_\text{CM}(\beta)$ is the translational partition function
of the center of mass \cite{StatMech00}:
\begin{align}
  Q_\text{CM}(\beta)
  =
  \frac{V}{\Lambda^3}
  \quad \text{with}\quad
  \Lambda = h\left(\frac{\beta}{2\pi M}\right)^{\frac{1}{2}}
  \label{eq:Qcm}
\end{align}
and $M$ is the total mass.
Since we use the Jacobi Hamiltonian, Eq.~(\ref{eq:Hamjac}), in the present work, 
we obtain to the rotational-vibrational partition
function (without the $Q_\text{CM}$ term).

%
%
\section{Path-integral formalism\label{ch:pi}}
\noindent%
Using the path-integral formalism of quantum mechanics and 
the Trotter factorization of an exponential function of two non-commutative 
(in this case the kinetic and the potential energy) operators, 
we write the quantum partition function in the form:
\begin{align}
  Q(\beta)
  &=
  \int \langle \rr{1} | e^{-\beta \hat{H}} | \rr{1} \rangle \: \dd \rr{1} \nonumber\\
  &=
  \int \langle \rr{1} | \left(e^{-\beta_M \hat{H}}\right)^M | \rr{1} \rangle \: \dd \rr{1} \nonumber\\
  &=
  \int 
  \langle \rr{1} | e^{-\beta_M \hat{H}}| \rr{2} \rangle
  \ldots
  \langle \rr{M-1} | e^{-\beta_M \hat{H}}| \rr{M} \rangle   
  \langle \rr{M} | e^{-\beta_M \hat{H}}| \rr{1} \rangle   
  \: \dd \rr{1} \ldots \dd \rr{M} \nonumber\\\\
  &\approx  
  \int \ldots \int
  \langle \rr{1} | e^{-\beta_M \hat{V}/2} e^{-\beta_M \hat{T}} e^{-\beta_M \hat{V}/2} | \rr{2} \rangle
  \ldots \nonumber \\
    &\quad\quad\quad\quad\quad
  \langle \rr{M-1} | e^{-\beta_M \hat{V}/2} e^{-\beta_M \hat{T}} e^{-\beta_M \hat{V}/2} | \rr{M} \rangle
  \ldots \nonumber \\
  &\quad\quad\quad\quad\quad
  \langle \rr{M} | e^{-\beta_M \hat{V}/2} e^{-\beta_M \hat{T}} e^{-\beta_M \hat{V}/2} | \rr{1} \rangle   
  \: \dd \rr{1} \ldots \dd \rr{M}, 
\end{align}
where $\beta_M=\beta/M$, $\rr{j}$ labels the coordinates of the the path-integral beads ($j=1,\ldots,M$), and $\hat{T}$ and $\hat{V}$ are the quantum mechanical kinetic and potential energy operators, respectively.
Using the properties of the position eigenfunctions,
the integrals can be evaluated in the usual way \cite{Cermp95,Tu02}
and if the kinetic energy operator has a simple form (no cross terms and constant coefficients),
we arrive to the well-known ring-polymer expression of the quantum partition function
\begin{align}
  Q_M(\beta)
  =
  &\left(\frac{1}{2\pi \betam}\right)^{fnM/2} 
  \left(\prod_{i=1}^{fn}m_i\right)^{M/2} \nonumber \\
  &\times\int\ldots\int\mr{d}\rr{1}\ldots\mr{d}\rr{M} %
  \exp\left[%
    -\betam%
    \sum_{j=1}^M
    \left\lbrace%
      \sum_{i=1}^{fn}
        \frac{1}{2}m_i\omem^2(\ri{j}{i}-\ri{j-1}{i})^2
      +
        V(\rr{j})
    \right\rbrace
  \right],
  \label{eq:qrp}
\end{align}
where the cyclic ``boundary condition'', $\bos{r}_{M+1}=\bos{r}_1$, is used 
to shorten the notation.
In Eq.~(\ref{eq:qrp}), 
$\omega_M=\beta_M^{-1}$ is the ring-polymer's harmonic angular frequency, and $fn$ 
is the total number of physical degrees of freedom.
For a molecule in the three-dimensional space $f=3$.
If the molecular Hamiltonian is written in LF coordinates (see eq.~(\ref{eq:Hamcart})),
$m_i$ is the physical mass for the $i=1,\ldots,n=N$ nucleus, while in the case of using Jacobi coordinates (see Eq.~(\ref{eq:Hamjac})),
the $m_i$ is replaced by the reduced mass, $\mu_i$ (see Appendix), 
and $n=N-1$ is the number of the Jacobi vectors.

The quantum partition function in Eq.~(\ref{eq:qrp}) can be considered as a classical configurational integral of an extended system, which includes
$M$ copies (beads) of the original system in which the neighboring beads are connected
with harmonic springs of $\omega_M=\beta_M^{-1}=M/\beta$ angular frequency.
In short, this is an $fnM$ dimensional hypothetical classical system at $\betam$ inverse temperature 
with the potential energy function
\begin{align}
  V_{\text{rp,tot}}(\rr{1},\ldots,\rr{M})
  =
    \sum_{j=1}^M
    \left\lbrace%
      \sum_{i=1}^{fn}
        \frac{1}{2}\mu_i\omem^2(\ri{j}{i}-\ri{j-1}{i})^2
      +
        V(\rr{j})
    \right\rbrace.
  \label{eq:rppes}
\end{align}
For the present discussion, it is an important difference from a truly classical system 
that the potential energy of the ring-polymerized system, $V_{\text{rp,tot}}$,
depends on the temperature through the angular frequency of the ring-polymer springs, $\omem=\betam^{-1}=M/\beta$.

%
%

\section{Nested-sampling integration}
\noindent%

The nested sampling technique had been originally introduced by Skilling \cite{Sk04,Sk06}, in the field of Bayesian probability and inference to sample high-dimensional spaces and computing Bayesian evidences,
\begin{equation}\label{pins_evidence1}
  \mathcal{Z}=\int \mathcal{L} (\theta) \pi (\theta)\: \dd\theta,
\end{equation}
where $\mathcal{L}(\theta)$ is a given likelihood and $\pi(\theta)$ a given prior probability density function.

The method was later adapted to be used to sample the potential energy surface of atomistic systems 
\cite{PaBaCs10,BaPaBa16,BaVa16,BaBeSa17,Pa18}, thus allowing the calculation of the partition function, 
which is normally inaccessible except for the simplest models.  

\begin{figure}
  \includegraphics[scale=0.4]{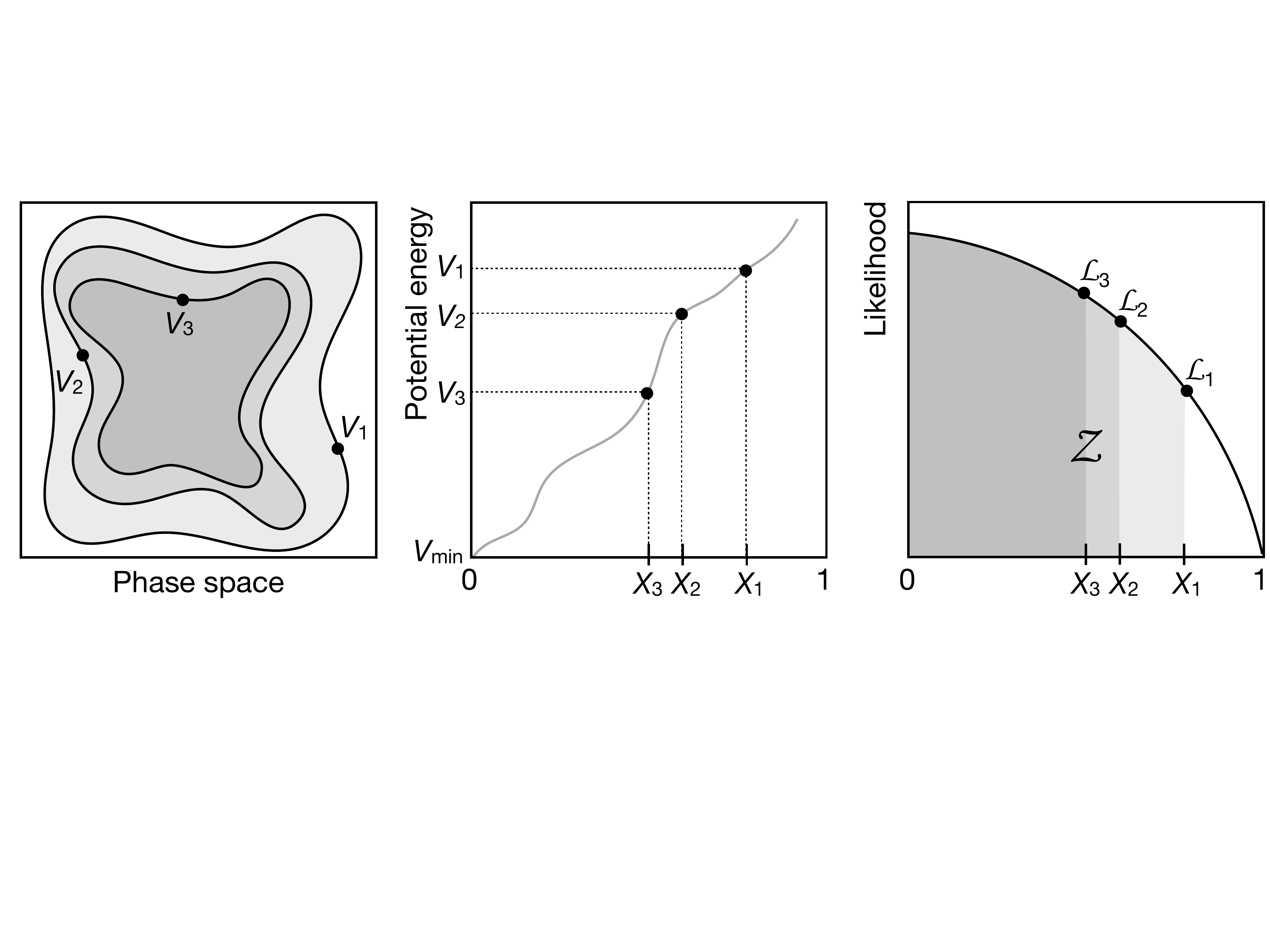} \\[-4cm]
  \caption{%
    Nested likelihood contours are sorted to enclosed prior mass $X$---likelihood function with area $\mathcal{Z}$.
    \label{fig:pins_int}
  }
\end{figure}

The cornerstone of the method is the reduction of the many-dimensional integral in Eq.~(\ref{pins_evidence1})
to a one-dimensional one. For this purpose, we define $X(\lambda)$
as the amount of prior mass with likelihood greater than some threshold $\lambda$,
\[
X(\lambda)=\int_{\mathcal{L}(\theta)>\lambda} \pi (\theta) \: \dd\theta,
\]
which transforms $\mathcal{Z}$ into the integral (see also Figure~\ref{fig:pins_int})
\begin{equation} \label{pins_evidence2}
  \mathcal{Z}=\int_{0}^{1} \mathcal{L}(X) \: \dd X.
\end{equation}

At the beginning of the sampling we take $\nns$ samples randomly from the prior $\pi$, this is going to be the initial sample set, then a series of iterations are performed.
At each iteration step, $i$, we choose the point with the lowest likelihood value, $\mathcal{L}_i$,
remove and replace it by a new point under the constraints that it has to be uniformly drawn from the phase space, where the likelihood is larger than
$\mathcal{L}_i$.
Now, one can approximate the evidence in Eq.~(\ref{pins_evidence2}) via a
quadrature sum over the removed points,
$
\mathcal{Z} \approx \sum_{i \in \textrm{removed points}} \mathcal{L}_i \left( X_i -X_{i+1} \right)
$.
Since the points are uniformly distributed, the value of the prior mass in the $i^\mathrm{th}$ iteration can be approximated as, $X_i \approx [\nns/(\nns+1)]^i$ \cite{Sk06}.
In terms of the atomistic PES, the likelihood is the Boltzmann factor, $\exp(-\beta V(\mx{r}))$, thus, 
nested sampling is a ``top-down" approach, starting from the high-energy region of the PES and going towards the global minimum through a series of nested energy levels. 
The iteration is stopped when the evidence is converged, i.e. the contribution by the latest likelihood value is smaller than a given pre-set tolerance~\cite{Sk06}. This stopping criterion highly depends on both the actual problem and the likelihood function, thus can be different for different purposes, however, in our case, terminating the sampling when $ \exp (-\beta V(\mx{r}_j)) X_j < 10^{-5}\mathcal{Z}$ is sufficient. The procedure of the method is also summarized in Figure~\ref{fig:nsproc}.


\begin{figure}[h!]
\begin{mdframed}[backgroundcolor=gray!20] 
\begin{enumerate}[itemsep=-1mm]
\item Take $\nns$ live points $\mx{r}_{(1)},\dots,\mx{r}_{(\nns)}$ uniformly from the entire PES,
and the corresponding likelihoods $\exp(-\beta V(\mx{r}_{(1)})),\dots,\exp(-\beta V(\mx{r}_{(\nns)}))$.
\item for $i=1,\dots,j$
 \begin{enumerate}[itemsep=-1mm]
  \item record the lowest likelihood $\exp (-\beta V(\mx{r}_i))$ 
  \item set $X_i=[\nns/(\nns+1)]^i$ and $w_i = X_{i-1}-X_{i}$,
  \item increment $\mathcal{Z}$ by $w_i \exp (-\beta V(\mx{r}_i))$, 
  \item replace $\exp (-\beta V(\mx{r}_i))$: draw a new sample $\mx{r}$ from the PES
  under the constraint that $\exp (-\beta V(\mx{r})) > \exp (-\beta V(\mx{r}_i))$.
  \end{enumerate}
\item Increment $\mathcal{Z}$ by $X_j \left[ \exp (-\beta V(\mx{r}_1))+...+\exp (-\beta V(\mx{r}_j)) \right]\nns^{-1}$.
\end{enumerate}
\end{mdframed}
\caption{Nested-sampling computational procedure\label{fig:nsproc}}
\end{figure}

\subsection{Computational strategy to generate a new sample point}
The initial samples are generated by randomly choosing $\nns$ points within the entire PES, \emph{i.e.,} generating configurations where 
the (Jacobi) position vectors are randomly placed within the simulation cell.  
Due the fast shrinkage of the phase space volume at lower energy levels, generating new configurations randomly in subsequent iterations quickly becomes impossible. 
To overcome this, new points are generated by cloning a randomly chosen existing sample and performing a random walk until a point is sufficiently independent from the parent configuration.
The random walk is a series of Monte Carlo steps, where the coordinates of the Jacobi position vectors are changed, 
with each step accepted if the Boltzmann-factor of the new configuration is lower than the current limit ($\exp(-\beta V_i(\mx{r}))$ in the $i^\mathrm{th}$ iteration).
The step size is adjusted throughout the sampling to have an acceptance ratio around 20\%. 

\subsection{Parallelization}
Similarly to earlier work on classical systems \cite{BaPaBa16,BaBeSa17}, we parallelize the nested sampling algorithm by evolving as many sample configurations as the number of processors ($n_p$) for an $n_w/n_p$ steps, instead of decorrelating the single cloned configuration alone for $n_w$ steps. Thus, each configuration will be evolved through $n_w$ steps on average before being recorded as the one with the lowest likelihood.

\subsection{Classical vs. quantum mechanical applications}

For classical interatomic potential models the energy function is independent of the temperature, thus the partition function can be calculated for arbitrary $\beta$ values from a single nested-sampling calculation \cite{PaBaCs10,PaBaCs14,BaPaBa16,BaBeSa17}. On the contrary, the PES of the ring-polymerized system is temperature dependent, Eq.~(\ref{eq:rppes}), 
so it is necessary to carry out independent nested sampling
iterations for every temperature value. However, the derivatives of the quantum partition function at a certain temperature could be computed with no additional cost, which is an aspect to be explored in a future work.

%
%
\section{Numerical results \label{ch:numres}}
\subsection{Test system: A nine-dimensional integral}
In order to test our implementation and check the applicability of nested sampling for 
integrals similar to those of the (ro)vibrational partition function of polyatomic molecules, we first computed 
\begin{align}
  \mathcal{I}_9
  &=
  \int\limits_{-\infty}^\infty\ldots\int\limits_{-\infty}^\infty 
    \text{exp}\left(-\sum\limits_{i=1}^9 \frac{1}{2} x_i^2\right)
    \dd x_1\ldots \dd x_9
  \\
  &=
  \left[%
   \int\limits_{-\infty}^\infty
     \text{exp}\left(-\frac{x_1^2}{2}\right)
     \dd x_1
  \right]^9 
  =
  \left[%
    \sqrt{2\pi}
  \right]^{9} \approx  3906.69
  \label{eq:testint}
\end{align}
as an explicit nine-dimensional integral.
Numerical results are shown in Table~\ref{tab:i9test}, calculated both with a relatively small ($\nns=1000$) and with a large ($\nns=20000$) sample. 
We shall use similar sample sizes in exploratory and
production runs, respectively, for molecular systems (see next section).

\begin{table}
  \caption{%
    Test calculation using our nested-sampling implementation for 
    the $\mathcal{I}_9$ integral defined in Eq.~(\ref{eq:testint}) and 
    evaluated as an explicit nine-dimensional integral.
    $\mathcal{I}_9=(2\pi)^{9/2}\approx 3906.69$.
    \label{tab:i9test}
  }
  \begin{tabular}{@{}c@{\ \ \ \ }c@{\ \ \ \ }c@{\ \ \ \ }c@{}}
    \hline\hline\\[-0.4cm]
    $\nns$$^\text{a}$ & 
    $\bar{\mathcal{I}}_9$$^\text{b}$ & 
    $\sigma_{\bar{\mathcal{I}}_9}$$^\text{b}$ & 
    $1-\bar{\mathcal{I}}_9/\mathcal{I}_9$ \\ 
    \hline\\[-0.4cm] 
     1000  & 3911.5   & 54.5  & 0.12\% \\
    20000  & 3903.6   & 18.9   & 0.08\% \\
    \hline\hline\\[-0.4cm] 
  \end{tabular}
  \begin{flushleft}
    $^\text{a}$ %
      Number of sample points. Each initial sample point was generated 
      from a uniform continuous probability density over $\left[-L/2,L/2\right]$ with $L=10$.
      New sample points were generated in an MCMC iteration including 800 steps and collective moves. \\
    $^\text{b}$ %
      $\bar{\mathcal{I}}_9$ and $\sigma_{\bar{\mathcal{I}}_9}$: 
      sample mean and sample variance of the mean 
      calculated from 20 independent runs.
  \end{flushleft}
\end{table}

\subsection{Rotating-vibrating molecules}
In order to demonstrate the applicability of the proposed
path-integral nested sampling method for real molecular systems, 
we have selected three polyatomic molecules of spectroscopic interest:  
the parent isotopologue of magnesium hydride, water, and ammonia.
For these, both accurate (and cost-efficient) potential energy surfaces as well as benchmark-quality quantum partition function values are already available in the literature. 

The rovibrational partition function values, calculated at several temperatures,
are collected in Table~\ref{tab:qpart}, and the number of sample points, $K$, the number of Markov chain Monte Carlo (MCMC) steps, $S$, and the number of beads, $M$, necessary for the current accuracy are also included. The number of nested-sampling iteration steps, $\niter$, was between $5\cdot 10^5$ and $1.5\cdot 10^6$ (generally the number of necessary iteration steps, $\niter$, increases linearly with the number of sample points, $K$). 

The overall computational cost of path-integral nested sampling is determined by the cost of a
single PES call multiplied with the number of PES calls, the latter being $M\times \niter\times K \times S$.
Both the number of sample points, $K$, and MCMC steps, $S$, has an effect on the accuracy, and as a rule of thumb, $K\times S$ determines the quality of the results (for sufficiently large $K$ and $S$ values)~\cite{BaBeSa17}.

Looking at the results it is remarkable, that path-integral nested-sampling can reproduce the Boltzmann sum
of the variational energies with a less than 1~\% relative error. 
At a given temperature, the larger the number of beads is, the more accurate the path-integral Trotter factorization becomes. 
However, at higher temperatures fewer beads are sufficient to achieve high accuracy, 
and thus, 
in the region where the variational technique becomes very expensive (or even unfeasible), 
path-integral nested sampling remains a practical alternative.

For a start, we performed exploratory computations for the three systems studied in this work using a small sample, $K=1000$, and only $S=1600$ MCMC steps. The sample mean of the path-integral nested-sampling results agreed with the variational reference values within a few percent, although the sample variance was large. These
exploratory calculations took a couple of minutes on a laptop for NH$_3$, for which we made use of the very fast polynomial PES of Ref.~\cite{YuBaTeThJe11}. The computations for the smaller, triatomic, H$_2$O molecule took longer than for NH$_3$ due to
the longer evaluation time of the water monomer PES 
(we used MB-pol PES implementation~\cite{MeBaPa13,BaMePa14}, which includes the monomer PES of Ref.~\cite{PaSc97}).
At the same time, we could use this NH$_3$ PES only up to ca. 3000~K. Beyond this temperature value a broader configuration space give significant contribution to the integral, Eq.~(\ref{eq:qrp}), for which unphysical regions of the polynomial PES hindered the computations.

To obtain the more accurate path-integral nested sampling results of Table~\ref{tab:qpart}, we used two larger parameter sets: one with $K=10^4$ sample points and $S=6.4\cdot 10^3$ MCMC steps, and a ca. four times larger one with $K=2\cdot 10^4$ and $S=1.2\cdot 10^4$. 
The sample mean values of the two computations are in excellent agreement with each other as well as with the variational reference values. We also note that by increasing $K$ and $S$ the sample variance of the mean is significantly reduced. 

In so far as it can be ascertained, the $\nns$ sample size---necessary to converge the integral---only slightly increases with the number of beads, and it has a very weak dependence on the number of atoms, $N$ (the ``real'' physical degrees of freedom). Based on nested-sampling studies of classical systems \cite{PaBaCs10,PaBaCs14,BaPaBa16,BaBeSa17}, we think that a larger $\nns$ value might be necessary for systems with a large number of important basins on the PES.
The efficiency of generating a new live point (now, determined by the number of MCMC steps, $S$) can perhaps be further improved by using total-energy Hamiltonian Monte Carlo \cite{BaBeSa17} instead of all-atom Monte Carlo moves used here, which is an aspect to be explored in future work.

\begin{table}
  \caption{%
    Rovibrational partition function values of selected molecular systems
    for a few representative temperatures:
    comparison of path-integral nested sampling (PI-NS) with variational results.
    [All $Q(T)$ values, including the classical ones, 
    correspond to an energy scale for which the zero coincides
    with the lowest vibrational energy eigenvalue, $E_0$. ] 
    \label{tab:qpart}
  }
\scalebox{1.}{%
  \begin{tabular}{@{} 
    c@{\ \ \ }
    c
    c@{\ \ \ }
    c@{\ \ \ }
    c
    c@{\ \ }c @{}}
    \hline\hline\\[-0.4cm]
    &
    \multicolumn{1}{c}{\raisebox{-0.5cm}{{Classical}$^\text{a}$}} & 
    \multicolumn{5}{c}{Quantum mechanical}  \\[-0.25cm]
    \cline{3-7}\\[-0.3cm]
    &
    \multicolumn{1}{c}{} & 
    Boltzmann sum$^\text{b}$ & 
    &
    \multicolumn{3}{c}{PI-NS$^\text{c}$} \\
    \cline{5-7}\\[-0.4cm]
    $T / \text{K}$  & 
    \multicolumn{1}{c}{$\bar{Q}_\text{cl}(\sigma_{\bar{Q}_{cl}})$} &
    $Q_\text{ref}/n_\text{SSW}$ & 
    &
    $M$  &
    \multicolumn{1}{c}{$\bar{Q}(\sigma_{\bar{Q}})$} &
    \multicolumn{1}{c}{$\bar{Q}(\sigma_{\bar{Q}})$} \\[0.2cm]
     & $K=2\cdot 10^4$ &  & & & $K=10^4$ & $K=2\cdot 10^4$ \\
     & $S=1.2\cdot 10^4$ &  & & & $S=6.4\cdot 10^3$ & $S=1.2\cdot 10^4$ \\
    \hline\\
      \multicolumn{7}{l}{$^{24}$MgH:$^\text{d}$} \\
    \hline\\[-0.4cm]      
    1000	&	155.7(6)&	142.4	&&	9&  143.0(1.5) & 142.1(1.7)	\\
    2000	&	401.1(1)&	404.1	&&	5&  402.4(4.7) & 403.4(3.2)	\\
    3000	&	833.1(3)&	804.8	&&	3& 809.9(28.8) & 807.4(6.2)	\\
    \hline\\
      \multicolumn{7}{l}{H$_2\ ^{16}$O:$^\text{e}$} \\    
    \hline\\[-0.4cm]      
    1000	&	5.84(3)$\cdot 10^3$&	6.09$\cdot 10^2$ && 12 & 6.01(36)$\cdot 10^2$  & 6.04(9)$\cdot 10^2$ \\
    2000	&	4.92(2)$\cdot 10^3$&	2.64$\cdot 10^3$ &&  6 & 2.62(11)$\cdot 10^3$  & 2.66(3)$\cdot 10^3$ \\
    3000	&	1.05(1)$\cdot 10^4$&	7.98$\cdot 10^3$ &&  3 & 8.04(13)$\cdot 10^3$  & 8.04(6)$\cdot 10^3$ \\
    4000	&	2.32(1)$\cdot 10^4$&	1.99$\cdot 10^4$ &&  3 & 1.98(2)$\cdot 10^4$  & 2.00(2)$\cdot 10^4$ \\
    \hline\\
      \multicolumn{7}{l}{NH$_3$:$^\text{f}$  } \\    
    \hline\\[-0.4cm]      
    1000	&	4.24(3)$\cdot 10^5$&	3.00$\cdot 10^3$ && 6	& 2.97(5)$\cdot 10^3$ & 3.02(4)$\cdot 10^3$ \\
    2000	&	1.62(1)$\cdot 10^5$&	2.94$\cdot 10^4$ && 4	& 2.97(8)$\cdot 10^4$ & 2.95(3)$\cdot 10^4$ \\
    3000	&	4.10(2)$\cdot 10^5$&	1.71$\cdot 10^5$ && 3	& 1.71(2)$\cdot 10^4$ & 1.72(2)$\cdot 10^5$ \\
    \hline
  \end{tabular}
}
  \begin{flushleft}  
  $^\text{a}$
  Classical partition function values were obtained with PI-NS using a single bead, $M=1$.\\
  $^\text{b}$
    Boltzmann sum of 
    variational rovibrational energy eigenvalues including 
    the appropriate degeneracy and
    spin-statistical weight factors. \\
  $^\text{c}$
    PI-NS: path-integral nested sampling. $M$ is the number of beads.
    The initial sample was generated from a uniform continuous probability density 
    function over $[-L/2,L/2]$.
    The sample mean, $\bar{Q}$, and 
    the sample variance of the mean, $\sigma_{\bar{Q}}$, were calculated from 
    20 independent runs.  \\
  $^\text{d}$
  We used the atomic masses $m(\text{Mg})=23.985\ 041\ 7$~u and $m(\text{H}) =1.007\ 825\ 0$~u
  and the PES taken from Refs.~\cite{ShHeRoBe07,SzCs15}.
  $L=8$; $n_\text{SSW}=1$. \\
  $^\text{e}$ 
  We used $m(\text{H})=1.007\ 276\ 5$, $m(\text{O})=15.990\ 526$; and 
  the PES of Ref.~\cite{MeBaPa13,BaMePa14}; 
  the $Q_\text{var}$ values were taken from Ref.~\cite{JPCRD16}; $L=6$;
  and $n_\text{SSW}=2$. \\
  $^\text{f}$
  We used $m(\text{N})=14.003\ 074$~u,
  $m(\text{H})=1.007\ 825\ 0$~u, and the PES of Ref.~\cite{YuBaTeThJe11}.
  the $Q_\text{var}$ values were taken from Ref.~\cite{SSHeYuHiTe14};
  $L=4$; $n_\text{SSW}=6$. \\
  \end{flushleft}
\end{table}

%
%
%
\clearpage
\section{Summary and conclusions\label{ch:sum}}
\noindent%
In this work we present the implementation and the first numerical applications
of a direct computational approach for the absolute rovibrational quantum partition function of small, polyatomic molecules of spectroscopic interest. The approach relies on the combination of the path-integral formalism of quantum mechanics and the nested sampling technique. Nested sampling is a multi-dimensional integration technique 
which can be efficiently used also for integrals which have regions which are exponentially localized but give significant contributions.
The computational cost of the path-integral nested-sampling method
is determined by the cost of a PES call, 
the number of important basins 
in the configuration space (the computation of fluxional systems is more demanding),
and it scales linearly with the number of the path-integral beads, 
the necessary number of which is approximately proportional with the inverse temperature.
Thereby, we expect path-integral nested sampling to be a feasible alternative to the 
Boltzmann summation technique of variationally computed energy levels 
for small, polyatomic systems at elevated temperatures,
and for larger systems for which the direct variational computation of 
a large number of eigenvalues is not possible. 

For the versatile applicability of the method, 
the development of new (or a boost of existing) accurate and cost-effective potential energy surfaces
is necessary. Since we anticipate the high-temperature region
(in which contributions also from unbound states is important)
particularly well suited for path-integral
nested sampling, the (accurate and cost-effective) PES should describe a sufficiently broad region of 
the configuration space to be able to account for high-temperature (and high-energy) phenomena,  
including dissociation.

%
%
\vspace{1cm}
\noindent %
\textbf{Acknowledgment} \\
B. Sz. and E. M. gratefully acknowledge the financial support 
of a PROMYS Grant (no. IZ11Z0\_166525)
of the Swiss National Science Foundation. 
B. Sz. also thanks the European Social Fund. EFOP-3.6.1-16-2016-0023.
During this work we used the computer cluster ATLASZ of ELTE and the NIIF Infrastructure in Debrecen. 
We thank Dr. Tamás Szidarovszky for sending to us his implementation of the $^{24}$MgH PES
based on \cite{ShHeRoBe07,SzCs15}.
L. B. P. acknowledges support from the Royal Society through a Dorothy Hodgkin Research Fellowship.
We also thank an Instant Access Grant of the ARCHER Supercomputing Center, which allowed us to test the scalability and applicability of the parallelized implementation for large-scale production runs.

\clearpage
\section*{Appendix: Definition of the Jacobi vectors, Jacobi Hamiltonian, and reduced masses}
\noindent%
The $N-1$ Jacobi vectors, $\mx{x}\in\mathbb{R}^{(N-1)\times 3}$, 
and the position vector of the center of mass, $\mx{R}_\text{CM}\in\mathbb{R}^3$ 
are constructed
in a linear transformation of the laboratory-frame (LF) Cartesian coordinates, 
$\bos{r}\in\mathbb{R}^{N\times 3}$ as
\begin{align}
  \left(%
  \begin{array}{@{}c@{}}
    \bos{x} \\
    \bos{R}_\text{CM} \\
  \end{array}
  \right)
  =
  (\bos{U}\otimes \bos{I}_3)
  \bos{r}, 
\end{align}
where the transformation matrix (see for example p.~10 of Ref.~\cite{SuVaBook98}) is 
\begin{align}
  \bos{U}
  =
  \left(%
  \begin{array}{@{}ccccc@{}}
    1 & -1 &  0 & \ldots & 0 \\
    \frac{m_1}{m_{12}} & \frac{m_2}{m_{12}} & -1 & \ldots & 0 \\
    \vdots & & & & \vdots \\
    \frac{m_1}{m_{12\cdots N-1}} &
    \frac{m_2}{m_{12\cdots N-1}} &     
    \ldots & \ldots & -1 \\ 
    \frac{m_1}{m_{12\cdots N}} &
    \frac{m_2}{m_{12\cdots N}} &     
    \ldots & \ldots & 
    \frac{m_N}{m_{12\cdots N}} \\     
  \end{array}
  \right)
\end{align}
with $m_{12\cdots n} = \sum_{i=1}^n m_i$.
Upon this linear transformation of the coordinates,  the  Jacobi determinant
is 1, and the Hamiltonian, 
Eq.~(\ref{eq:Hamcart}), transforms to
\begin{align}
  \hat{H}' 
  = 
  -\sum_{i=1}^{N-1}\sum_{\alpha} 
    \frac{1}{2\mu_i} \frac{\partial^2}{\partial x^2_{i\alpha}}
  -\sum_{\alpha} 
    \frac{1}{2M} \frac{\partial^2}{\partial R^2_{\text{CM},\alpha}}
  +V(\mx{x}).\label{eq:transform}
\end{align}
where the reduced masses are
\begin{align}
  \frac{1}{\mu_i} 
  =
  \frac{1}{m_{i+1}} + \frac{1}{m_{1\cdots i}}
\end{align}
and $M=m_{12\cdots N}$ is the total mass.
After subtracting the kinetic energy operator of the center of mass, 
the Jacobi Hamiltonian is obtained as
\begin{align}
  \hat{H}
  = 
  -\sum_{i=1}^{N-1}\sum_{\alpha} 
    \frac{1}{2\mu_i} \frac{\partial^2}{\partial x^2_{i\alpha}}
  +V(\mx{x}).\label{eq:transform2}
\end{align}

\providecommand{\latin}[1]{#1}
\providecommand*\mcitethebibliography{\thebibliography}
\csname @ifundefined\endcsname{endmcitethebibliography}
  {\let\endmcitethebibliography\endthebibliography}{}

\clearpage
\textbf{ ``For Table of Contents only'' } \\
\includegraphics[scale=0.6,angle=90]{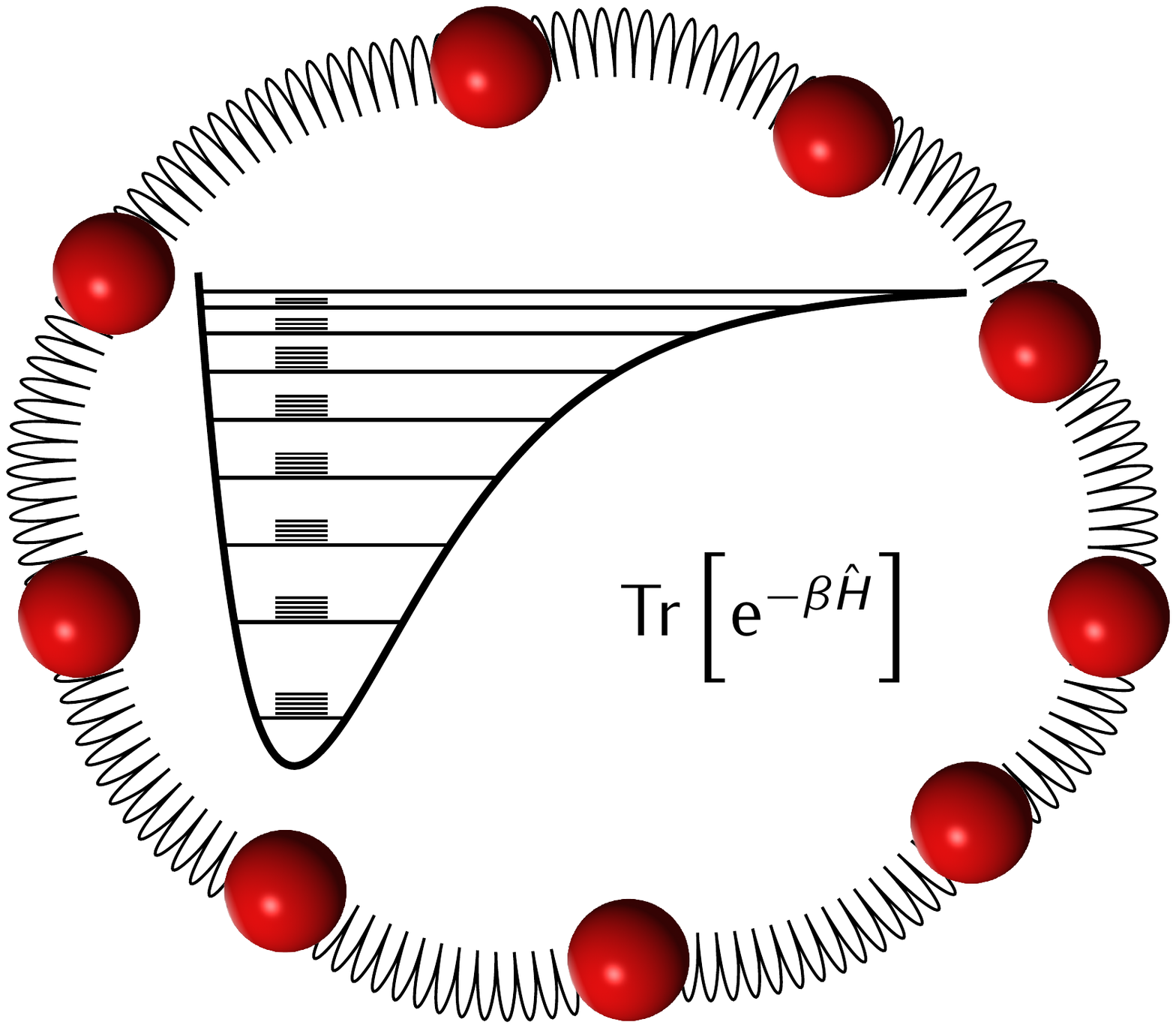}\quad\quad\quad\quad\quad

\end{document}